\documentclass{elsart}
\usepackage{graphicx}
\usepackage{amsmath}
\usepackage{amssymb}
\begin{document}

\begin{frontmatter}

\title{Quantum Criticality and Superconductivity in Quasi-Two-Dimensional Dirac Electronic Systems}

\author{E. C. Marino} and
\author{Lizardo H. C. M. Nunes}

\address{Instituto de F\'{\i}sica,  Universidade Federal do Rio de Janeiro,\\
Caixa Postal 68528,  Rio de Janeiro-RJ 21941-972, Brasil}

\begin{abstract}
We present a theory describing the superconducting (SC) interaction
of Dirac electrons in a quasi-two-dimensional system consisting of a
stack of $N$ planes. The occurrence of a SC phase is investigated
both at $T=0$ and $T\neq 0$, in the case of a local interaction,
when the theory must be renormalized and also in the situation where
a natural physical cutoff is present in the system. In both cases,
at $T=0$, we find a quantum phase transition connecting the normal
and SC phases at a certain critical coupling. The phase structure is
shown to be robust against quantum fluctuations. The SC gap is
determined for $T=0$ and $T\neq 0$, both with and without a physical
cutoff and the interplay between the gap and the SC order parameter
is discussed. Our theory qualitatively reproduces the SC phase
transition occurring in the underdoped regime of the high-Tc
cuprates. This fact points to the possible relevance of Dirac
electrons in the mechanism of high-Tc superconductivity.
\end{abstract}

\begin{keyword}
Dirac electrons, superconductivity, quantum criticality
\end{keyword}

\end{frontmatter}

\section{Introduction}\label{int}

Surprisingly, there are many condensed matter systems in one and two
spatial dimensions containing electrons that may be described by a
relativistic, Dirac-type lagrangian, namely Dirac electrons. Even
though these are evidently non-relativistic systems, this fact
occurs because in some materials there are special points in the
Brillouin zone where two bands touch in a single point around which
the electron dispersion relation may be linearized as $\epsilon(\vec
k) = v_{ \rm F} |\vec k|$. The kinematics of these electrons can be
described by a Dirac-type lagrangian where the velocity $v_{ \rm F}$
determines the angle of the Dirac cone. At the tip of this cone the
Fermi surface reduces to a point, the Fermi point, and the density
of states vanishes. The elementary excitations around a Dirac point
are Dirac electrons. They are, after all, a result of the
electron-lattice interaction.

There are many important quasi-two-dimensional systems containing
Dirac electrons. Among them we could mention the high-Tc cuprates,
where Dirac points appear in the intersection of the nodes of the
d-wave superconducting gap and the 2D-Fermi surface. Because of
these nodes the low-energy quasiparticle spectrum is gapless and the
dispersion relation can be linearized as described above
\cite{slee}. The low-energy quasiparticle dynamics is determined
exclusively by these points, since they are occupied even at very
low temperatures \cite{ferr,herbut,ft,and}. For these reasons, Dirac
electrons are expected to play an important role in the cuprates.
This motivates the application of the model introduced below to
these materials and the results thereby obtained seem to point
towards this direction.

Dirac electrons also appear in semi-metals such as graphene sheets
or stacks thereof, namely graphite, where the vanishing density of
states at the Fermi points has important consequences in the
electronic properties like, for instance, the absence of screening
of the Coulomb potential \cite{sem,hald,guinea,cn,zhang,nov}.

Another class of materials where the presence of Dirac electrons
produces interesting effects are the rare-earth dichalcogenides such
as $2H-TaSe_2,\ 2H-NbSe_2,\  2H-TaS_2 $ and $2H-NbS_2$. In these
systems charge-density-wave order coexists with superconductivity at
low temperatures and Dirac points form in the intersection of the
Fermi surface with the nodal lines of the charge-density-wave order
parameter \cite{castroneto,cnuch}. This is a particular example of
nodal liquids, which in general contain Dirac electrons \cite{bfn}.
Finally, we may list carbon nanotubes as another important type of
materials that have also been shown to possess Dirac electrons
\cite{bf}.

In this work, we present a theory describing the superconducting
interaction of Dirac electrons associated to two distinct Dirac
points belonging to a stack of $N$ planes. We analyze the conditions
for the existence of a superconducting gap, both at $T=0$ and $T\neq
0$, either for finite $N$ or in the limit $N \rightarrow\infty$,
which corresponds to the case when the system is actually
three-dimensional.

At $T=0$, we show that the system presents a quantum critical point
separating the normal and superconducting phases and determine the
superconducting gap as a function of the coupling constant. The
theory is renormalized in a $1/N$ expansion. A renormalization group
analysis is then performed, demonstrating the independence of
physical quantities from the renormalization point. We also
investigate the effect of quantum fluctuations in our results and
demonstrate that the phase diagram found in mean field is robust
against these fluctuations.

We then consider the finite temperature case and determine the
superconducting gap $\Delta$ as a function of the temperature and of
the zero temperature gap. We find a critical temperature $T_c$,
where the gap vanishes. The possibility of occurrence of dynamical
generation of a superconducting gap without the corresponding
spontaneous breakdown of the U(1) symmetry, in compliance with the
Coleman-Mermin-Wagner-Hohenberg \cite{mw} theorem is discussed in
detail, as well as the associated Kosterlitz-Thouless transition
suffered by the phase of the complex order parameter.

We finally make a detailed study of the situation usually found in
condensed matter applications, where a natural momentum cutoff
exists in the system, both at $T=0$ and $T \neq 0$. We consider the
general regime, where the cutoff is not necessarily much larger than
the gap, as well as the weak coupling regime where the cutoff is
much larger than the gap. The former situation is likely to be
relevant for high-Tc superconductors while the latter is the one
found in conventional BCS superconductors.

The quantum phase transition occurring in our model and the behavior
of $T_c$ around the quantum critical point qualitatively reproduce
very well the superconducting transition in the high-Tc cuprates in
the underdoped region. This suggests that Dirac electrons may play
an important role in the mechanism of high-Tc superconductivity.

\section{The Model and the Effective Action}\label{mod}

We consider a quasi-two-dimensional electronic system consisting of
a stack of planes containing two Dirac points. In addition, we
introduce an internal index $a=1,...,N$, supposed to characterize
the different planes to which the electrons may belong. The electron
creation operator, therefore, is given by $\psi^\dag_{i\sigma a}$,
where $i=1,2$ are the Dirac indices, corresponding to the two Fermi
points, $\sigma = \uparrow,\downarrow$, specifies the z-component of
the electron spin and $a=1,...,N$ labels the electron plane. We
assume, further, that there is a BCS-type superconducting
interaction, whose origin is understood to be determined by some
underlying microscopic theory. In the case of the high-Tc cuprates,
in particular, there is no consensus about what would be such a
theory, however it is generally accepted that superconductivity is
constrained to the $CuO_2$ planes.

The complete lagrangian we will consider is given by
\begin{equation}
\mathcal{ L }  ={\rm i} \overline\psi_{ \sigma a} \not\! \partial \
\psi_{ \sigma a} + g \left (\psi^\dag_{1\uparrow a} \
\psi^\dag_{2\downarrow a} + \psi^\dag_{2\uparrow a} \
\psi^\dag_{1\downarrow a}  \right ) \left (\psi_{2\downarrow b} \
\psi_{1\uparrow b} + \psi_{1\downarrow b} \ \psi_{2\uparrow b}
\right ), \label{L}
\end{equation}
where $g > 0$ is a constant that may depend on some external control
parameter, such as the pressure or the concentration of some dopant.
Later on, we will make $g \equiv \frac{\lambda}{N}$.

 We use the
following convention for the Dirac matrices:
\begin{equation}
\gamma^0 = \sigma^z, \ \ \gamma^0\gamma^1 = \sigma^x, \ \
\gamma^0\gamma^2 = \sigma^y , \label{gammas}
\end{equation}

Observe that the interaction lagrangian contains four terms,
describing the various possible BCS interactions in which a Cooper
pair would form between electrons with opposite spins, belonging to
different Fermi points but in the same plane. Nevertheless, the
interaction of electrons belonging to different planes is allowed.

Furthermore, we can motivate our model out of a microscopic
description of the high-Tc cuprates by noting that t-J model
calculations of the electron spectral function indicate the
emergence of small pockets of Dirac electrons at low doping
\cite{wl}.

In addition to a complex valued O(N) symmetry \cite{mm}, the
lagrangian above possesses the U(1) symmetry
\begin{equation}
 \psi_{i\sigma a}
\rightarrow e^{{\rm i}\theta} \psi_{i\sigma a} \ \ \ \ \ \ \ \ \ \
\psi^\dag_{i\sigma a} \rightarrow e^{-{\rm i}\theta}
\psi^\dag_{i\sigma a}\ \ \ \ \ \ \ \ \ \   i=1,2 \label{u1}
\end{equation}
and the chiral U(1) symmetry
\begin{equation}
 \psi_{1\sigma a}
\rightarrow e^{{\rm i}\theta} \psi_{1\sigma a} \ \ \ \ \ \ \ \ \ \
\psi_{2\sigma a} \rightarrow e^{-{\rm i}\theta} \psi_{2\sigma a} .
\label{u2}
\end{equation}

We shall see below that the former symmetry is spontaneously broken
at $T=0$. A model presenting the spontaneous breakdown of the latter
has been studied in \cite{semenoff}.

We now introduce a Hubbard-Stratonovitch complex scalar field
$\sigma$ through
\begin{eqnarray}
\mathcal{ Z }_{ \sigma }
& = &
\int \mathcal{ D } \sigma^{ * } \mathcal{ D }
\sigma \exp \left \{ - {\rm i}
\int d^3 x \frac{ 1 }{ g }\ \sigma^{* } \sigma \right \}
\nonumber
\\
& = &
\int \mathcal{ D } \sigma^{ * } \mathcal{ D }
\sigma \exp\left\{- {\rm i}
\int d^3 x \frac{ 1 }{ g }
\left[
\sigma^{ * } - g\
\left(\psi^\dag_{1\uparrow a} \
\psi^\dag_{2\downarrow a} + \psi^\dag_{2\uparrow a} \
\psi^\dag_{1\downarrow a}
\right)
\right]
\right .
\nonumber
\\
& &
\hspace{4.5cm}
\left .
 \times
 \left[ \sigma - g\ \left(
\psi_{2\downarrow b} \ \psi_{1\uparrow b} + \psi_{1\downarrow b} \
\psi_{2\uparrow b}\right) \right ] \right \} .
\label{zsigma1}
\end{eqnarray}
In terms of this, we may write the partition function
\begin{equation}
\mathcal{ Z } = \frac{ 1 }{ \mathcal{ Z}^{ 0 } _{ \Psi } } \int
\mathcal{ D } \Psi^{ \dagger }_{ \sigma } \mathcal{ D } \Psi_{
\sigma } \exp\left\{ {\rm i }\int d^3 x \mathcal{L } \right\}
\end{equation}
as
\begin{equation}
\mathcal{ Z } = \frac{ 1 }{ \mathcal{ Z }_{ \sigma } \mathcal{ Z }^{
0 } _{ \Psi } } \int \mathcal{ D }  \Psi^{ \dagger } \mathcal{ D }
\Psi \mathcal{ D } \sigma^{ * } \mathcal{ D } \sigma \ e^{ {\rm i }
\int d^{ 3 } x \mathcal{ L } \left[ \Psi , \sigma \right] } ,
\end{equation}
where
\begin{eqnarray}
\mathcal{ L } \left[ \Psi, \sigma \right]
& = &
{\rm i} \overline\psi_{ \sigma a} \not\! \partial \ \psi_{ \sigma a}
- \frac{1}{ g }\ \sigma^{ * }  \sigma
- \sigma^{ * } \left( \psi_{2\downarrow b} \ \psi_{1\uparrow b}
+ \psi_{1\downarrow b} \ \psi_{2\uparrow b}\right)
\nonumber
\\
& &
- \sigma \left(\psi^\dag_{1\uparrow a} \
\psi^\dag_{2\downarrow a} + \psi^\dag_{2\uparrow a} \
\psi^\dag_{1\downarrow a} \right).
\label{lsigpsi}
\end{eqnarray}
>From this we obtain the field equation for the auxiliary field
$\sigma$:
\begin{equation}
\sigma = - g\ \left (\psi_{2\downarrow a} \ \psi_{1\uparrow a} +
\psi_{1\downarrow a}\ \psi_{2\uparrow a} \right ) \label{eqsigma}
\end{equation}

As we shall see, the vacuum expectation value of $\sigma$ will be
taken as the order parameter for the superconducting phase.

We will now integrate over the fermionic fields. In
order to do that, we introduce the Nambu fermion field $\Phi^\dag_a
= (\psi^\dag_{1\uparrow a}\ \psi^\dag_{2\uparrow a}\
\psi^\dag_{1\downarrow a}\ \psi^\dag_{2\downarrow a} )$. In terms of
this we can rewrite (\ref{lsigpsi}) as
\begin{equation}
\mathcal{ L } \left[ \Psi, \sigma \right] =  - \frac{1}{ g }\
\sigma^{ * }  \sigma \ +\ \Phi^\dag_a \mathcal{ A } \Phi_a
,\label{lsigpsi1}
\end{equation}
where the matrix $\mathcal{ A }$ is given, in momentum space, by
\begin{equation}
\mathcal{ A} =
\begin{pmatrix}
-k_0 & k_{-} & 0 & -\sigma \\
k_{+} & -k_0 & -\sigma & 0 \\
0 & -\sigma^* & -k_0 & - k_{+} \\
- \sigma^* & 0 & - k_{-} & -k_0
\end{pmatrix}
\end{equation}
with $k_{\pm} = v_{\rm F} \left (k_2 \pm \rm{i} k_1 \right )$.

Integrating on the fermion fields and redefining the coupling constant as $g =
\frac{\lambda}{N}$, we obtain
\begin{equation}
\mathcal{ Z } = \frac{ 1 }{ \mathcal{ Z }_{ \sigma } } \int
\mathcal{ D } \sigma^{ * } \mathcal{ D } \sigma \ e^{ {\rm i }
S_{eff} \left[
 \sigma \right] } \label{zfinal},
\end{equation}
where
\begin{equation}
S_{eff} \left[ \sigma \right]  = \int d^{ 3 } x \left( -
\frac{N}{\lambda} |\sigma |^2
 \right) - {\rm i} N \ln {\rm Det}\left [
\frac{\mathcal{ A} \left[ \sigma \right]}{\mathcal{ A} \left[ 0
\right]} \right].
\end{equation}

The determinant of the matrix $\mathcal{ A}$ is $\ \det \mathcal{
A}[\sigma] = \left [(k_0^2 - v^2_{\rm F}|\vec k|^2) - |\sigma |^2
\right ]^2$, hence the above expression becomes
\begin{equation}
S_{eff} \left[ \sigma \right]  = \int d^{ 3 } x \left( -
\frac{N}{\lambda} |\sigma |^2
 \right) - {\rm i}2 N \rm{Tr} \ln \left [
1 + \frac{|\sigma |^2}{\Box}\right] \label{seff}
\end{equation}

\section{The Superconducting Transition at $T=0$ and Quantum Criticality}\label{t0qc}

Let us consider in this section the $T=0$ case. We shall see that a
quantum phase transition occurs, connecting the superconducting and
normal phases.

\subsection{The Renormalized Effective Potential}

 Since we are considering the zero temperature case, the
functional integral in (\ref{zfinal}) must be dominated by constant
configurations of $\sigma$, which minimize the effective potential
per plane, $V_{\rm eff}$, corresponding to (\ref{seff}). This is
more conveniently evaluated in the euclidean space and is given by
\begin{equation}
V_{\rm eff}\left( |\sigma| \right) = \frac{  |\sigma|^{ 2 } }{
\lambda } - 2 \int \frac{ d^2 k }{ \left( 2 \pi \right)^2 } \int
\frac{ d \omega }{ 2 \pi } \left\{ \ln \left[ 1+ \frac {|\sigma|^2}{
\omega^2 + v^2_{ \rm{F} } k^2} \right] \right \},
 \label{veff}
\end{equation}
where, henceforth, by  $\sigma$ we actually mean $ \langle 0|\sigma
|0\rangle$. The above expression corresponds to a mean field
approximation. Conversely, this would be the leading order
approximation in an $1/N$ expansion and would be the exact result
for $N\rightarrow \infty$.

The explicit form of the effective potential may be evaluated in
this framework, by introducing a large momentum cutoff  $\Lambda/
v_{ \rm{F}}$. The resulting expression, obtained from (\ref{veff})
is
\begin{equation}
V_{ \rm{eff} } \left( |\sigma| \right)= \frac{|\sigma|^2}{ \lambda}
- \frac{\Lambda\ |\sigma|^2}{2\pi v^2_{ \rm{F} }}   + \frac{
|\sigma|^{3}}{3\pi v^2_{ \rm{F} }} .
 \label{veff1}
\end{equation}

Quartic fermionic theories in 2+1D have been shown to be
renormalizable in a $1/N$ expansion \cite{ros}. In order to
eliminate the divergent constant $\Lambda$ from (\ref{veff1}) we
renormalize the coupling constant $\lambda$ using the usual
renormalization condition \cite{coleman}
\begin{equation}
\left. \frac{
\partial^2
V_{ {\rm eff } } }{
\partial \sigma
\partial \sigma^*
} \right|_{ |\sigma| = \sigma_0 } = \frac{ 1 }{ \lambda_{ R },
}\label{lr}
\end{equation}
where $\sigma_0$ is an arbitrary finite scale parameter, the
renormalization point and $ \lambda_{ R }$ is the (finite),
renormalized coupling constant.

Inserting (\ref{veff1}) in (\ref{lr}), we obtain
\begin{equation}
\frac{ 1 }{ \lambda_R } = \frac{ 1 }{ \lambda } - \frac{ \Lambda }{
\alpha } + \frac{ 3 \sigma_0 }{ 2 \alpha }  ,\label{lambdaren}
\end{equation}
where $\alpha \equiv 2\pi  v^2_{ \rm{F} }$.

Substituting this result in (\ref{veff1}), we get the renormalized
effective potential per plane
\begin{equation}
V_{ { \rm eff},R } \left( |\sigma| \right) = \frac{ |\sigma|^2 }{
\lambda_R } - \frac{ 3 \sigma_0 }{ 2 \alpha } |\sigma|^2 + \frac{
2}{ 3 \alpha } |\sigma|^3 . \label{veffr}
\end{equation}

\subsection{The Gap Equation and the Quantum Critical Point}

Let us study now the minima of the renormalized effective potential
per plane, Eq. (\ref{veffr}). The first and second derivatives of
$V_{\rm eff,R}$ with respect to $|\sigma|$ are given, respectively,
by
\begin{equation}
V'_{ { \rm eff},R } \left( |\sigma| \right) = 2 |\sigma| \left(
\frac{ 1 }{ \lambda_R } - \frac{ 3 \sigma_0 }{ 2 \alpha }  + \frac{
|\sigma| }{ \alpha }\right) \label{v1}
\end{equation}
and
\begin{equation} V''_{ { \rm eff},R } \left( |\sigma|  \right) = 2
 \left( \frac{ 1 }{ \lambda_R } - \frac{ 3 \sigma_0 }{ 2\alpha }
 + \frac{ 2|\sigma| }{ \alpha }\right)\label{v2}
\end{equation}

The ground state is determined by the solutions of $V'_{ \rm{eff},R
} = 0$. This admits two solutions, which we call $\Delta$. Notice
that
\begin{equation}
\Delta = \left|\langle 0|\sigma |0\rangle\right| \label{v22}
\end{equation}
and $\langle 0|\sigma |0\rangle$ is a complex order parameter for
superconductivity.

>From (\ref{v1}) we conclude that either $\Delta =0$ or $\Delta
\neq 0$, the nonzero solutions satisfying the gap equation
\begin{equation}
1 = \frac{\lambda_R}{\alpha} \left( \frac {3\sigma_0}{2}  - \Delta
\right ). \label{gap}
\end{equation}
Inserting the $\Delta =0$ solution in (\ref{v2}), we get
\begin{equation}
V''_{ \rm{eff},R } \left(\Delta =0 \right) = 2 \left( \frac{ 1 }{
\lambda_R } - \frac{ 3 \sigma_0 }{ 2\alpha } \right) \label{der2}
\end{equation}
and we conclude that $\Delta =0$ will be a minimum only for
$\lambda_R <  2\alpha / 3\sigma_0 \equiv \lambda_c $.

>From (\ref{gap}), conversely, we see that the $\Delta \neq 0$
solution is given by
\begin{equation}
\Delta_0 = \alpha\left(\frac{1}{\lambda_c} - \frac{  1  }{ \lambda_R
}\right) \label{delta0}.
\end{equation}

Since $\Delta_0$ is positive semi-definite, the above expression
will actually be a solution only for $\lambda_R > \lambda_c$. On the
other hand, from (\ref{v2}), we immediately see that $V''_{
\rm{eff},R } \left(  \Delta \right)= 2\Delta_0 / \alpha > 0$
for the solutions of the gap equation (\ref{gap}).

We can infer that the ground state of the system will be
\begin{equation}
 \Delta_0 =
 \left \{  \begin{array}{c}
 0 \ \ \ \ \ \ \ \ \ \ \ \ \
 \lambda_R < \lambda_c \\    \\
 \alpha\left(\frac{1}{\lambda_c} - \frac{  1  }{ \lambda_R
}\right)
  \ \ \ \ \ \ \ \ \  \lambda_R > \lambda_c
       \end{array} \right .
 \label{s00}.
\end{equation}

Expression (\ref{s00}) implies that the system undergoes a
continuous quantum phase transition at the quantum critical point
$\lambda_c =  4 \pi v^2_{ \rm{F} } / 3 \sigma_0 $. Since
$\Delta$ is the modulus of the order parameter for
superconductivity, we conclude that, for couplings below $\lambda_c$
the electronic system will be in the normal state, while for
couplings above $\lambda_c$, it will be in a superconducting one
(for temperature effects, see teh next section). The quantum
critical point $\lambda_c$, therefore, separates a normal from a
superconducting phase.

\subsection{Renormalization Group Analysis}

Let us now apply renormalization group methods, in order to show
that our results are completely independent of the arbitrary finite
scale $\sigma_0$ introduced in our renormalization procedure.

We start showing that the renormalized effective potential $V_{\rm
 eff,R}$, given by (\ref{veffr}) satisfies a renormalization group
 equation. Indeed it is easy to see that
\begin{equation}
\left( \sigma_0 \frac{
\partial
}{
\partial \sigma_0
} + \beta \frac{
\partial
}{
\partial \lambda_R
} \right) V_{ {\rm eff },R } = 0
\end{equation}
with the $\beta$-function given by $\beta = -  3 \sigma_0 / 2
\alpha \, \lambda^2_R = - \lambda^2_R / \lambda_c $. This means
that the renormalized effective potential does not depend on the
renormalization point $\sigma_0$. A negative $\beta$-function, on
the other hand, implies that the theory is asymptotically free.
Indeed, keeping $\Delta_0$ fixed and taking the limit $\sigma_0
\rightarrow\infty$ we see that $\lambda_R \rightarrow 0$.

One can also show that the gap $\Delta_0$, given by (\ref{delta0})
satisfies the same renormalization group equation being, therefore,
also independent of $\sigma_0$. Indeed, using (\ref{delta0}) and the
expression of the $\beta$-function just found, we obtain
\begin{equation}
\left( \sigma_0 \frac{
\partial
}{
\partial \sigma_0
} + \beta \frac{
\partial
}{
\partial \lambda_R
} \right)  \Delta_0  = 0.
\end{equation}

Finally, solving the differential equation corresponding to the
$\beta$-function definition, namely,
\begin{equation}
\sigma_0 \frac{
\partial \lambda_R
}{
\partial \sigma_0
} = \beta ,
\end{equation}
we obtain
\begin{equation}
 \frac{1}{ \lambda_R (\sigma'_0)} -  \frac{1}{ \lambda_c (\sigma'_0)}
=  \frac{1}{ \lambda_R (\sigma''_0)} -  \frac{1}{
\lambda_c(\sigma''_0)} ,
\end{equation}
where $\frac{1}{ \lambda_c (\sigma'_0)} = \frac{3
\sigma'_0}{2\alpha}$ and $\sigma'_0$ and $\sigma''_0$ are two
arbitrary values of the renormalization point. The above equation,
clearly shows that the superconducting gap $\Delta_0$, given by
(\ref{delta0}) is independent of the scale $\sigma_0$.

The theory predicts the existence of a quantum critical point
$\lambda_c$, separating two phases at $T=0$: a normal ($\Delta_0=0$)
and a superconducting one ($\Delta_0 \neq 0$). Nevertheless, the
theory does not predict the value of $\lambda_c$. This has to be
determined experimentally. The renormalization group analysis,
however, guarantees that the physics of the quantum phase transition
will not depend on the renormalization point $\sigma_0$.

\subsection{Gaussian Quantum Fluctuations}

Let us now investigate whether the results we found in the
saddle-point approximation for the ground state of the fermionic
system described by (\ref{L}) are robust against quantum
fluctuations at $T=0$. Notice that for the actual existence of the
quantum phase transition it is necessary that the normal phase found
in the saddle point approximation should not be removed by higher
order corrections, as it happens, for instance, in the
Coleman-Weinberg mechanism \cite{coleman}. For this it is crucial
that the corrected gap equation admits a $\Delta_0 = 0$ solution.

Expanding the effective action (\ref{seff}) in (\ref{zfinal}) about
a stationary point $\sigma$ and retaining the gaussian quantum
fluctuations $\eta$, we obtain, after integrating over $\eta$ and
$\eta^*$,
\begin{equation}
\tilde{ S}_{ \rm{eff} } \left[ \sigma \right] = S_{ \rm{eff} }
\left[ \sigma \right] - {\rm Tr}\ln \mathcal{M}[\sigma],
 \label{seffl}
\end{equation}
where $S_{ \rm{eff} } \left[ \sigma \right]$ is given by
(\ref{seff}) and
\begin{equation}
\mathcal{M} =
\begin{pmatrix}
\frac{\delta^2 S}{\delta\sigma\delta\sigma^* }& \frac{\delta^2
S}{\delta\sigma^2 }\\
\frac{\delta^2 S}{\delta\sigma^{*2} } & \frac{\delta^2
S}{\delta\sigma^*\delta\sigma }
\end{pmatrix}[\sigma]
\end{equation}

The effective potential corresponding to this is given by
\begin{equation}
\tilde {V}_{ \rm{eff} } \left( |\sigma| \right)= V_{ \rm{eff} }
\left( |\sigma| \right) + {\mathcal V}\left( |\sigma| \right),
 \label{veffl}
\end{equation}
where $V_{ \rm{eff} } \left( |\sigma| \right)$ is given by
(\ref{veff1}) and
\begin{eqnarray}
{\mathcal V} \left( |\sigma| \right)
& = &
- \int \frac{ d^2 k }{ \left( 2 \pi \right)^2 } \int \frac{ d \omega }{ 2 \pi }
\left\{ \ln \left[ \frac{1}{\lambda}- a ( \omega, \vec k) \right]
\right.
\nonumber
\\
& &
\hspace{3.2cm}
\left.
+ \ln \left[ \frac{1}{\lambda} - a ( \omega, \vec k)
+ 2 |\sigma|^2 b ( \omega, \vec k) \right]
\right \},
 \label{veffl1}
\end{eqnarray}
with
\begin{equation}
a ( \omega, \vec k, |\sigma|) =  \int \frac{ d^2 q }{ \left( 2 \pi
\right)^2 } \int \frac{ d \theta }{ 2 \pi } \frac{2}{\left[v_{
\rm{F} }|\vec q| + {\rm i} \theta \right ]  \left[v_{ \rm{F} }|\vec
q +\vec k| - {\rm i} \left(\theta + \omega\right)\right ] +
|\sigma|^2}
\end{equation}
and $b ( \omega, \vec k, |\sigma|) = - \frac{\partial\ a}{\partial
|\sigma|^2}$.

We would like to stress at this point that, for the expansion about
the saddle-point to make sense, we must have ${\mathcal V}\left(
\sigma \right) \ll V_{ \rm{eff} } \left( \sigma \right)$. This
condition is actually usually satisfied because the expansion is a
power series in $\hbar$. Keeping this fact in mind, let us look for
the minima of $\tilde {V}_{ \rm{eff} } \left( \sigma \right)$. From
(\ref{veffl1}), we can infer that the first derivative of ${\mathcal
V}\left( \sigma \right)$ is of the form
\begin{equation}
{\mathcal V}\ '\left(| \sigma| \right) = |\sigma|\ f(|\sigma|),
\label{mvl}
\end{equation}
where $f(0)$ is a finite constant. Combining (\ref{mvl}) with
(\ref{v1}), we immediately conclude that $\Delta_0 = 0$ is a
solution of the corrected gap equation $\tilde {V}'_{ \rm{eff} }
\left( \sigma \right) =0$. The second derivative of the corrected
effective potential $\tilde {V}_{ \rm{eff} } \left( \sigma \right)
$, evaluated at this point is
\begin{equation}
\tilde {V}''_{ \rm{eff} } \left( \Delta =0 \right) = 2 \left( \frac{
1 }{ \lambda_R } - \frac{ 3 \sigma_0 }{ 2\alpha } \right)
 + {\mathcal V}\
''\left( \Delta_0 =0 \right), \label{mvll}
\end{equation}
where we have used (\ref{der2}). Since $\frac{{\mathcal V}\
''}{V''_{ \rm{eff} }} \propto \hbar^2 $, it follows that the second
term in the rhs of the above equation must be much smaller than the
first one. Therefore the sign of $\tilde {V}''_{ \rm{eff} } \left(
\sigma =0 \right)$ must be the same as the one of ${V}''_{ \rm{eff}
} \left( \sigma =0 \right)$. We conclude that $\Delta_0 =0$ is
indeed a minimum, even considering the quantum fluctuations. For the
same reason, the mean field solution $\Delta_0 \neq 0$ should not be
cancelled by quantum corrections. Thus, we conclude that the phase
structure we found at $T=0$ is robust against quantum fluctuations.

\section{ The Superconducting Transition at $T\neq 0$}

\subsection{The Gap Equation at $T \neq 0$}

We consider in this section the finite temperature effects in the
superconducting transition and in the order parameter $\Delta$. The
nonzero solutions for $\Delta$ at a finite temperature are supposed
to hold {\it a priori} only in the $N\rightarrow\infty$ limit,
because otherwise they are ruled out by the
Coleman-Mermin-Wagner-Hohenberg theorem \cite{mw}. This limit
corresponds to a physical situation where the three-dimensionality
of the system is explicitly taken in account. For finite values of
$N$ and $T\neq 0$, the situation is quite subtle. We discuss it in
the next subsection.

At $T \neq 0$, the effective potential is no longer given by
(\ref{veff}). It must be replaced by
\begin{equation}
V_{\rm eff} \left( |\sigma|, T \right) = \frac{ |\sigma|^2 }{
\lambda } - 2T \int \frac{ d^2 k }{ \left( 2 \pi \right)^2 } \sum_{
n =  - \infty }^{ \infty } \left\{ \ln \left[ 1 +
\frac{|\sigma|^2}{\omega^2_n + v^2_{ \rm{F} }  k^2} \right] \right\}
\label{veftemp},
\end{equation}
where $\omega_n = (2 n+1)\pi T$ are the fermionic Matsubara
frequencies corresponding to the functional integration over the
electron fields.

 The finite temperature corrections do not alter the ultraviolet
 divergence structure of the theory, hence we may eliminate the
 divergences in the $T\neq 0$ case through the same renormalization
 of the coupling constant $\lambda$ as in the zero temperature
 case, given by (\ref{lambdaren}).

In order to derive the gap equation, we consider the following
condition, which must be satisfied by the order parameter at a
finite temperature:
\begin{equation}
V'_{ \rm{eff} } \left( |\sigma|, T \right) = 2 |\sigma| \left\{
\frac{ 1 }{ \lambda } - \frac{ 1}{ 2\alpha } \int_0^{ \Lambda^2 }d x
\frac{ 1 }{ \sqrt{ x + |\sigma|^2 } } \tanh\left( \frac{ \sqrt{ x +
|\sigma|^2 } }{ 2 T } \right) \right\} =0 .
\end{equation}
This was obtained by taking the derivative of (\ref{veftemp}) with
respect to $|\sigma|$ and performing the Matsubara sum. In the above
equation, $\Lambda$ is the same cutoff used before.

As in the $T=0$ case, this admits two solutions, either
$\Delta(T)=0$ or $\Delta(T) \neq 0$. In the latter case, the
superconducting order parameter satisfies the gap equation
\begin{equation}
1 = \frac{  \lambda}{ \alpha} \int_{\Delta}^{\Lambda } d y
\tanh\left( \frac{ y }{ 2 T } \right)\label{gapeq}.
\end{equation}

\begin{figure}[ht]
\centerline {
\includegraphics
[clip,width=0.9\textwidth,angle=90
] {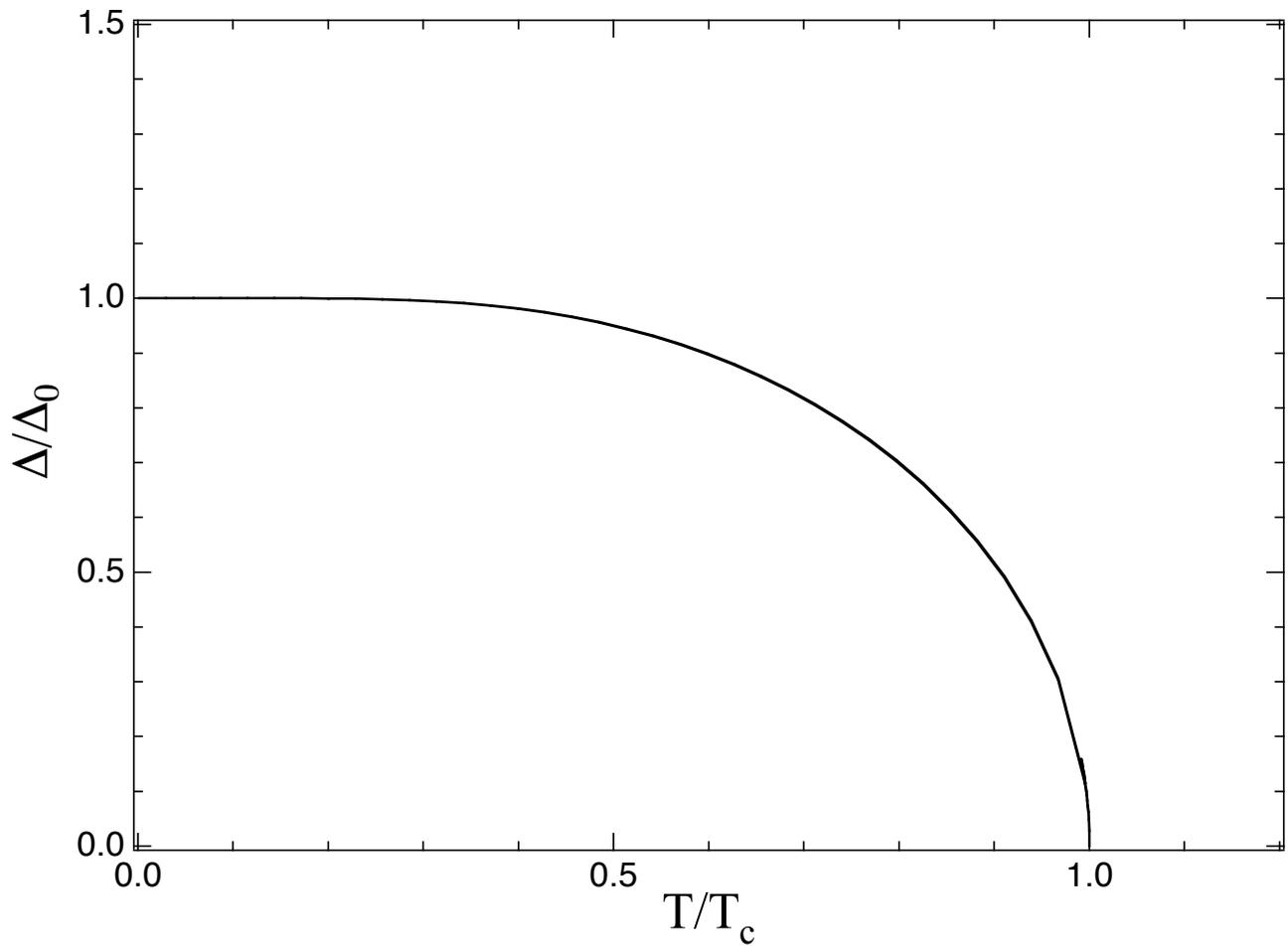} } \caption{The normalized superconducting gap
$ \Delta / \Delta_0 $ as a function of the normalized temperature $
T / T_c $.} \label{fig1}
\end{figure}

Solving the integral and renormalizing the coupling $\lambda$ as in
(\ref{lambdaren}), we find a general expression for the
superconducting gap as a function of the temperature, namely
\begin{equation}
\Delta (T) = 2 T \cosh^{- 1}\left[ \frac{e^{\frac{\Delta_0}{2T}}}{2}
\right ], \label{gap1}
\end{equation}
where $\Delta_0$ is given by (\ref{s00}), see Fig. \ref{fig1}. From
(\ref{gap1}) we can verify that indeed $\Delta (T=0)=\Delta_0$. Also
from the above equation, we may determine the critical temperature
$T_c$ for which the superconducting gap vanishes. Using the fact
that $\Delta (T_c)= 0$, we readily find from (\ref{gap1})
\begin{equation}
T_c =  \frac {\Delta_0}{2 \ln 2} \label{gap3}.
\end{equation}
This relation has been previously found in systems with a natural
cutoff in the weak coupling regime \cite{castroneto} (see next
section). It has also appeared in systems with dynamical mass
generation in 2+1D \cite{oko,rwp,appel,babaev}.

In Fig. \ref{fig2}, using (\ref{s00}) and (\ref{gap3}), we display
$T_c$ as a function of the coupling constant. This qualitatively
reproduces the superconducting phase transition of the high-Tc
cuprates in the underdoped region. Since our theory describes the
generic superconducting interaction of two-dimensional Dirac
electrons, we may see this result as an indication of the possible
relevance of this type of electrons in the high-Tc mechanism.

 In terms of the critical temperature, we may also express the gap as
\begin{equation}
\Delta (T) = 2 T \cosh^{- 1}\left[
2^{\left(\frac{T_c}{T}-1\right)}\right ] \label{gap4} \ .
\end{equation}
Near $T_c$, this yields
\begin{equation}
\Delta (T) \stackrel{T\lesssim T_c}{\sim} 2 \sqrt{2\ln 2}\ T_c\left
(1- \frac{T}{T_c} \right)^{\frac{1}{2}}\label{gap5} \ ,
\end{equation}
which presents the typical mean field critical exponent $1/2$.

We would like to remark, finally, that both the gap $\Delta(T)$ and
the critical temperature do not depend on the arbitrary
renormalization point $\sigma_0$.

\subsection{Dynamical Gap Generation versus Spontaneous Symmetry Breaking}

The well-known Coleman-Mermin-Wagner-Hohenberg theorem  \cite{mw}
forbids the occurrence of spontaneous breakdown of a continuous
symmetry at a nonzero temperature for systems in two spatial
dimensions. Our superconducting order parameter is complex and given
by
\begin{equation}
 \langle 0|\sigma |0 \rangle = \Delta  e^{{\rm i} \theta}
 \label{gap04} \ ,
\end{equation}
where $\Delta$ is the gap.
 Given the form of the field $\sigma$, (\ref{eqsigma}), we
infer from (\ref{gap04}) that a nonzero value for the gap would
imply, in principle, the spontaneous breakdown of the U(1) symmetry,
(\ref{u1}). For $N\rightarrow\infty$, the system is effectively
three-dimensional and the occurrence of a nonzero gap $\Delta (T)$
as determined in this section is is not in conflict with the
theorem. For $T=0$, either for finite or infinite $N$, also we have
a non-vanishing superconducting gap, which was studied in section 3.
In both cases the Coleman-Mermin-Wagner-Hohenberg theorem does not
apply and a nonzero gap leads to a nonzero order parameter according
to (\ref{gap04}).

In the case of finite $N$ and $T\neq 0$, in order to comply with the
theorem and yet having a superconducting phase, we may invoke the
mechanism proposed by Witten \cite{witt}, by means of which we may
have dynamical generation of a superconducting gap without the
corresponding U(1) symmetry breakdown. It goes as follows. Whenever
the gap is nonzero, according to (\ref{gap04}), we must shift the
field $\sigma$ as
\begin{equation}
\sigma\rightarrow\sigma - \Delta e^{{\rm i} \theta}
 \label{gap005} \
\end{equation}
in (\ref{lsigpsi}). This will produce an extra term in the effective
lagrangian (\ref{lsigpsi}). In terms of new fermion fields, defined
as
\begin{equation}
\hat{\psi}_{i\sigma a} \equiv e^{{- \rm i} \frac{\theta}{2}}\
\psi_{i\sigma a} \label{gap07},
\end{equation}
the extra term in (\ref{lsigpsi}) reads
\begin{equation}
\Delta\left[ \left(\hat{\psi}^\dag_{1\uparrow a} \
\hat{\psi}^\dag_{2\downarrow a} + \hat{\psi}^\dag_{2\uparrow a} \
\hat{\psi}^\dag_{1\downarrow a} \right) +  \left(
\hat{\psi}_{2\downarrow b} \ \hat{\psi}_{1\uparrow b} +
\hat{\psi}_{1\downarrow b} \ \hat{\psi}_{2\uparrow b}\right)\right]
 \label{gap06}.
\end{equation}
This is an explicit superconducting gap term that will make
\begin{equation}
\langle 0|\hat{\psi}^\dag_{1\uparrow a} \
\hat{\psi}^\dag_{2\downarrow a} + \hat{\psi}^\dag_{2\uparrow a} \
\hat{\psi}^\dag_{1\downarrow a} |0 \rangle \neq 0.
\end{equation}
Since the U(1) symmetry acts as $\psi_{i\sigma a} \rightarrow
e^{{\rm i}\omega} \psi_{i\sigma a}$ and $\theta\rightarrow\theta +
2\omega$ we immediately see that the field $\hat{\psi}_{i\sigma a}$
is invariant under U(1) rotations and therefore the non-vanishing
expectation value above does not imply spontaneous breakdown of the
U(1) symmetry (notice that the chiral U(1) symmetry (\ref{u2}) is
also unbroken). Thus, we can have dynamical generation of a
superconducting gap without the associated spontaneous symmetry
breaking \cite{witt}.

We must examine the thermodynamic conditions for the occurrence of
this situation. This has been done in detail for the case of the
Gross-Neveu model in 2+1D \cite{babaev} and also in the case of the
semimetal-excitonic insulator transition that occurs in layered
materials \cite{kv}, both related to the potential spontaneous
breakdown of the chiral symmetry. The results of these analysis also
apply here.

The basic point is that, in order to check whether the order
parameter (\ref{gap04}) is zero or not, we must analyze the
thermodynamics of the phase $\theta$ of the superconducting order
parameter. It turns out that this phase decouples and suffers a
Kosterlitz-Thouless \cite{kt} transition at a temperature $T_{KT}$.
For temperatures above $T_{KT}$ there is no phase coherence and the
superconducting order parameter vanishes because then $\langle
\cos\theta\rangle =\langle \sin\theta\rangle=0$ (even though
$\Delta$ may be different from zero). Below $T_{KT}$ there is a
phase ordering and there will be a nonvanishing gap provided the
condition $T<T_c$ is also met (otherwise $\Delta=0$). As it is,
$T_{KT} \leq T_c$ \cite{babaev} and, therefore, the actual
superconducting transition occurs at $T_{KT}$. It can be shown that
$T_{KT} \stackrel{N\rightarrow\infty}{\longrightarrow} T_c$
\cite{babaev}. This clearly indicates that in spite of the fact that
we may have a superconducting gap at a finite temperature in
two-dimensional space, only in a really three-dimensional system we
will have phase coherence developing at the same time that the
modulus of the order parameter becomes nonzero, as determined by the
gap equations.

It has been speculated \cite{pg} that the above scenario could
provide a framework for explaining the pseudogap transition that
precedes the superconducting transition in high-Tc cuprates in the
underdoped region. Our model provides a concrete realization of this
mechanism.

\section{Systems with a Physical Cutoff $\Lambda$}

\subsection{Physical Cutoff}

When considering applications in condensed matter systems, one
usually finds a natural energy cutoff $\Lambda$ (momentum cutoff
$\Lambda/v_{ \rm{F}}$). The Debye frequency (energy) is an example,
in the case of conventional BCS superconductors. In this case, no
renormalization is needed and the coupling constant $\lambda$ is the
physical one. We investigate in this section the modifications that
will occur in the superconducting electronic system under
consideration when there is a physical cutoff in energy or momentum.

The two-body interaction in this case, instead of being a delta
function leading to the local interaction in (\ref{L}), is given, in
terms of the momentum cutoff, by
\begin{equation}
 V(\vec k) =
 \left \{  \begin{array}{c}
 -\ \lambda\ \ \ \ \ \ \ \ \
 |\vec k| < \Lambda/ v_{ \rm{F} } \\    \\
 0 \ \ \ \ \ \ \ \ \  |\vec k| > \Lambda/ v_{ \rm{F} }
       \end{array} \right .
 \label{vk}.
\end{equation}
In coordinate space, this corresponds to the interaction potential
\begin{equation}
V(\vec r) = -\ \lambda \ \frac{\Lambda}{2\pi v_{ \rm{F} }|\vec r|}\
J_1\left(\frac{\Lambda |\vec r|}{2\pi v_{ \rm{F} } }\right)
 \label{vr},
\end{equation}
where $J_1$ is a Bessel function.

\subsection{The $T=0$ Case}

We must now evaluate (\ref{veff}) with a finite physical momentum
cutoff $\Lambda/v_{ \rm{F}}$. This yields
\begin{equation}
V_{ \rm{eff} } \left( |\sigma| \right)= \frac{|\sigma|^2}{ \lambda}
-\frac{2}{3\alpha}\left [ \left( |\sigma|^2 + \Lambda^2 \right
)^{\frac{1}{2}} - |\sigma|^3 - \Lambda^3 \right]
 \label{veff11}
\end{equation}
We would like to stress that we are not assuming that $\Lambda$ is
large compared to $|\sigma|$, rather, we are considering a
completely arbitrary finite cutoff $\Lambda$. This is not the
situation usually found in conventional BCS superconductors.
However, it is likely to be found in nonconventional ones such as
high-Tc cuprates. For $\Lambda \gg |\sigma|$ (\ref{veff11}) would
reproduce (\ref{veff1}).

The solutions of
\begin{equation}
V'_{ \rm{eff} } \left( |\sigma| \right)= 2|\sigma| \left[ \frac{1}{
\lambda}- \frac{\left( |\sigma|^2 + \Lambda^2 \right
)^{\frac{1}{2}}}{\alpha} +  \frac{|\sigma|}{\alpha}\right] =0
\label{veff22}
\end{equation}
will give the gap in the present case. This admits two solutions,
namely, $\tilde{\Delta}_0=0$ or
\begin{equation}
\tilde{\Delta}_0 = \frac{\lambda\alpha}{2}\left[
\frac{\Lambda^2}{\alpha^2} - \frac{1}{\lambda^2} \right].
 \label{g1}
\end{equation}

The second derivative of the potential, evaluated at
$\tilde{\Delta}_0=0$ is
\begin{equation}
V''_{ \rm{eff} } \left( \tilde{\Delta}_0=0 \right)=
\frac{1}{\lambda} - \frac{\Lambda}{\alpha} \label{veff33}
\end{equation}
and we conclude that $\tilde{\Delta}_0=0$ is a solution only for
$\lambda < \tilde{\lambda}_c$, with $\tilde{\lambda}_c =
\alpha / \Lambda $.  Conversely, the second derivative of
(\ref{veff11}) evaluated at $\tilde{\Delta}_0 \neq 0$ ( given by
(\ref{g1})) is positive for $\lambda > \tilde{\lambda}_c$. As a
consequence the gap now will be given by
\begin{equation}
 \tilde{\Delta}_0=
 \left \{  \begin{array}{c}
 0 \ \ \ \ \ \ \ \ \ \ \ \ \
 \lambda < \tilde{\lambda}_c \\    \\
 \frac{\alpha\lambda}{2}\left(\frac{1}{\tilde{\lambda}^2_c} - \frac{  1  }{
 \lambda^2
}\right)
  \ \ \ \ \ \ \ \ \  \lambda > \tilde{\lambda}_c
       \end{array} \right . ,
 \label{s000}
\end{equation}
which should be compared with (\ref{s00}), see Fig. \ref{fig3}.
Again we find a quantum phase transition, now at the critical
coupling $\tilde{\lambda}_c =  \alpha / \Lambda $. Observe now
that, contrary to the local case, since $\Lambda$ is a physical
parameter, the value of the quantum critical point
$\tilde{\lambda}_c$ is predicted by the theory.

Observe that, for $\lambda > \tilde{\lambda}_c$
\begin{equation}
\tilde{\Delta}_0 = \alpha\left(  \frac{1}{\tilde{\lambda}_c} -
\frac{  1  }{
 \lambda}  \right) \left(  \frac{\lambda +\tilde{\lambda}_c}{2 \tilde{\lambda}_c} \right)
 \label{g2}
\end{equation}
and in the region where $\lambda \gtrsim \tilde{\lambda}_c$, we have
$\tilde{\Delta}_0 \simeq \Delta_0$, with $\Delta_0$ given by
(\ref{s00}). This coincides with the regime where $\Lambda \gg
\tilde{\Delta}_0$ and occurs near the quantum critical point. Also,
from (\ref{g2}) we see that the quantum phase transition is in the
same universality class as the one found in Sect. 2, as it should.

\subsection{The $T \neq 0$ Case}

Let us consider now the case of systems with a physical cutoff at a
finite temperature. In this case the gap equation (\ref{gapeq}) must
be replaced by
\begin{equation}
1 = \frac{  \lambda}{ \alpha}
\int_{\tilde{\Delta}}^{\left(\tilde{\Delta}^2 +\Lambda^2
\right)^{\frac{1}{2}} } d y \tanh\left( \frac{ y }{ 2 T }
\right)\label{gapeq}.
\end{equation}

Solving the integral, we find an implicit equation for the
superconducting gap in the presence of a physical cutoff $\Lambda$,
at an arbitrary temperature, namely
\begin{equation}
\tilde{\Delta} (T) = 2 T \cosh^{- 1}\left[ e^{-
\frac{\alpha}{2T\lambda}}
 \cosh \left[ \frac{\left(\tilde{\Delta}^2(T) +\Lambda^2
\right)^{\frac{1}{2}}}{2T}\right]\right ] \label{gap44} \ .
\end{equation}

Using the fact that $\tilde{\Delta}_0$ satisfies (\ref{veff22}), we
may verify that indeed
\begin{equation}
\tilde{\Delta} (T) \stackrel{T\rightarrow 0}{\thicksim} 2 T \cosh^{-
1}\left[ \frac{e^{ \frac{\tilde{\Delta}_0}{2T}}}{2}
 \right] \stackrel{T\rightarrow 0}{\longrightarrow}\tilde{\Delta}_0
 \label{gap55},
\end{equation}
where $\tilde{\Delta}_0$ is given by (\ref{s000}).

Let us now determine the critical temperature $T_c$, for the onset
of superconductivity in the presence of a physical cutoff. We must
have $\tilde{\Delta} (T_c)=0$ and therefore, from (\ref{gap44})
\begin{equation}
 \cosh \left( \frac{\Lambda}{2T_c}\right)=  e^{\frac{\alpha}{2T_c\lambda}}\label{gap66} \ .
\end{equation}

This equation yields the following relation between $T_c$ and the
zero temperature gap, (\ref{s000})
\begin{equation}
\tilde{\Delta}_0 = \left( \frac{\lambda +
\tilde{\lambda}_c}{2\tilde{\lambda}_c}\right )2T_c \ln \left[
\frac{2}{1+ e^{-\frac{\Lambda}{T_c}}}\right]\label{gap77},
\end{equation}
where $\tilde{\lambda}_c=\alpha/\Lambda$.

A particular regime that is frequently studied is the one where the
cutoff is large compared to the critical temperature and to
$\tilde{\Delta}_0$, namely, when $\Lambda \gg T_c$ and
$\lambda\gtrsim \tilde{\lambda}_c$. Observe that this last condition
guarantees that $\Lambda \gg \tilde{\Delta}_0$, according to
(\ref{s000}) and (\ref{g2}). Since the previous relations imply
\begin{equation}
T_c \ll \Lambda = \frac{\alpha}{\tilde{\lambda}_c} \simeq
\frac{\alpha}{\lambda} \label{wc},
\end{equation}
we may infer that the conditions above hold in the weak coupling
regime. In this case, (\ref{gap77}) becomes simply $\tilde{\Delta}_0
= 2 \ln 2 \ T_c  $ \cite{castroneto}. Also, according to
(\ref{s000}) and (\ref{g2}) $\tilde{\Delta}_0 \rightarrow \Delta_0$
in this regime. Thus we recover (\ref{gap3}), the relation
previously found between $T_c$ and the zero temperature gap in
systems without a natural cutoff . This relation gives the ratio
$ \tilde{\Delta}_0 / T_c \sim 1.39$, which should be compared
with the corresponding ratio in the BCS theory for conventional
superconductors, namely, $1.76$, which is also derived in the limit
$ \Lambda / T_c  \gg 1$.

In the weak coupling regime, given by (\ref{wc}), we also recover
expressions (\ref{gap4}) and (\ref{gap5}), for the superconducting
gap as a function of the temperature. The pre-factor in the latter
expression, describing the behavior of the gap around $T_c$ is 2.36,
whereas the corresponding value in BCS theory is 3.06.

We solve (\ref{gap44}) numerically for $\tilde{\Delta }$, using
different values for the ratio $\lambda/\tilde{\lambda}_c$ and
display the result in Fig. \ref{fig4}. In this, we may observe that
indeed in the weak coupling regime the ratio $\Delta_0 / T_c $
approaches the result given by (\ref{gap3}). On the other hand, as
the coupling increases, we see that this ratio surpasses the value
obtained in BCS theory and approaches the values obtained in
strongly coupled systems.

There are condensed matter systems for which the weak coupling
condition (\ref{wc}) is not valid. Should one of such systems
contain Dirac fermions, we should use (\ref{s000}) and (\ref{gap44})
for the superconducting gap, respectively at $T=0$ and $T \neq 0$.
The critical temperature, by its turn, would be given by
(\ref{gap77}).

It is remarkable that the expression we find for the superconducting
gap in the case of Dirac fermions, strongly differs from the one
obtained in BCS theory in any dimensions. There the gap has an
exponential dependence on the inverse of the coupling $\lambda$.
Here, in spite of still being non-analytical in the coupling
constant $\lambda$, the gap has a power-law dependence on it and
vanishes below a critical value at zero temperature. The different
behavior can be traced back to the fact that in the case of Dirac
fermions the density of states vanishes at the Fermi points. In the
BCS case, however, we have a Fermi surface with a non-vanishing
density of states at the Fermi level and the momentum integrals used
for obtaining the effective potential may be evaluated as
\begin{equation}
\int \frac{d^2k}{(2\pi)^2} \simeq N(\epsilon_{ \rm{F} })
\int_{-\Lambda}^{\Lambda} d \xi \label{aprox},
\end{equation}
where $N(\epsilon_{ \rm{F} })$ is the density of states at the Fermi
level. This leads to a gap proportional to
$e^{-\frac{1}{N(\epsilon_{ \rm{F} })\lambda}}$.


\section{Conclusions}\label{last}

\begin{figure}[ht]
\centerline {
\includegraphics
[clip,width=0.9\textwidth,angle=90
] {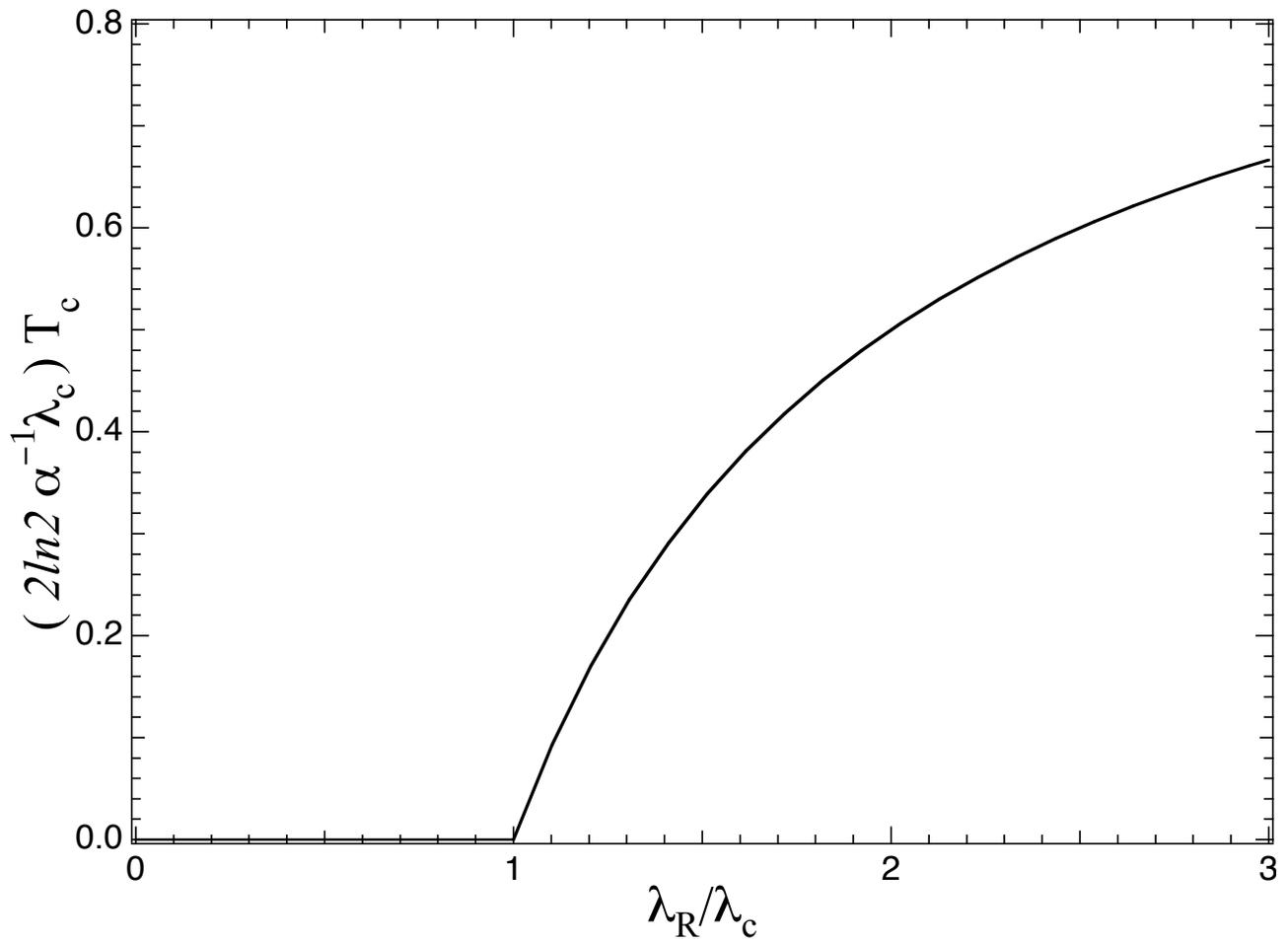} } \caption{The superconducting critical
temperature $ T_c $ as a function of the renormalized coupling  $
\lambda_R  $.} \label{fig2}
\end{figure}

\begin{figure}[ht]
\centerline {
\includegraphics
[clip,width=0.9\textwidth,angle=90
] {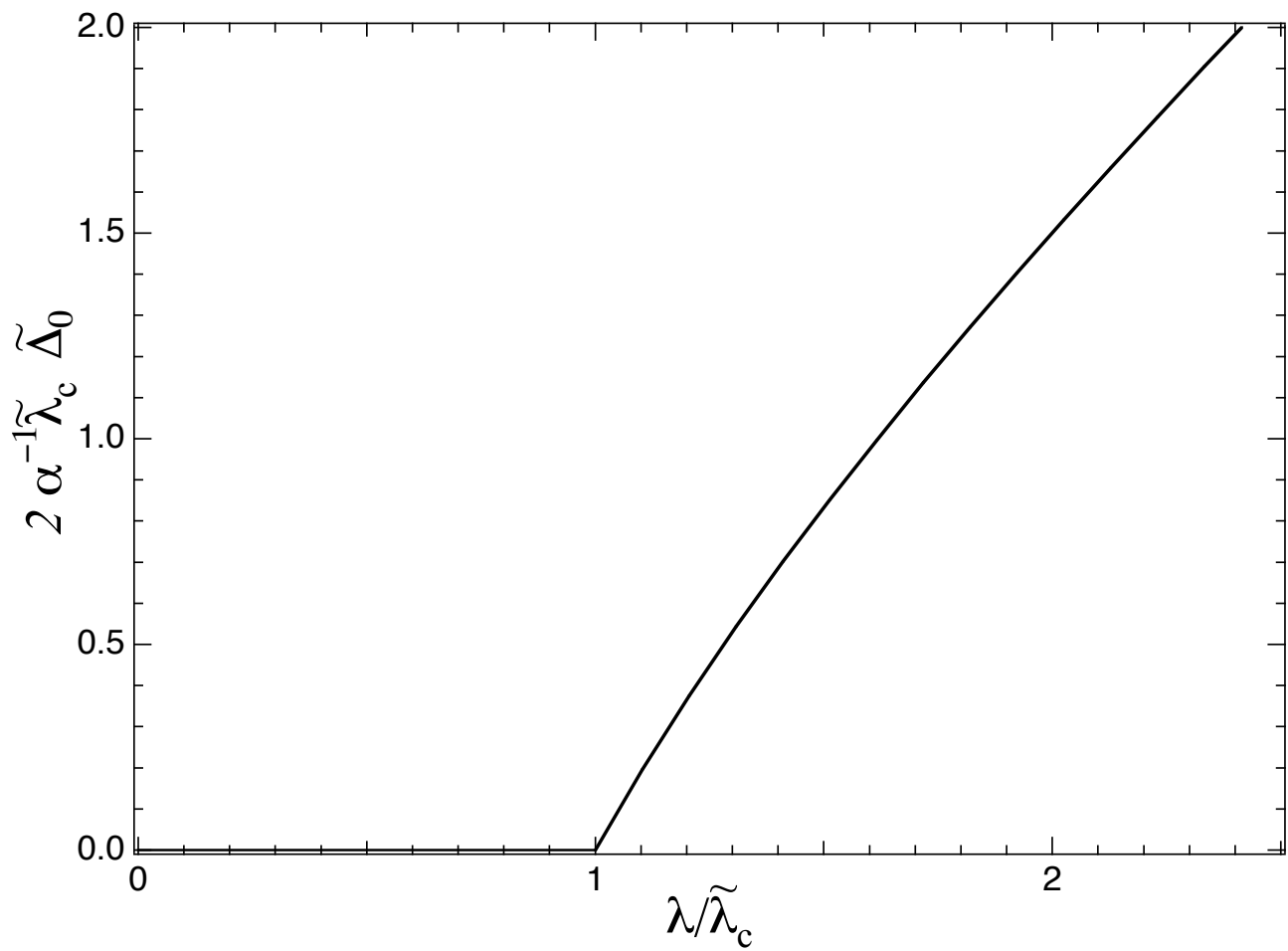} } \caption{ The zero temperature
superconducting gap as a function of the normalized coupling $
\lambda / \tilde{\lambda}_c $. } \label{fig3}
\end{figure}

\begin{figure}[ht]
\centerline {
\includegraphics
[clip,width=0.9\textwidth,angle=90
] {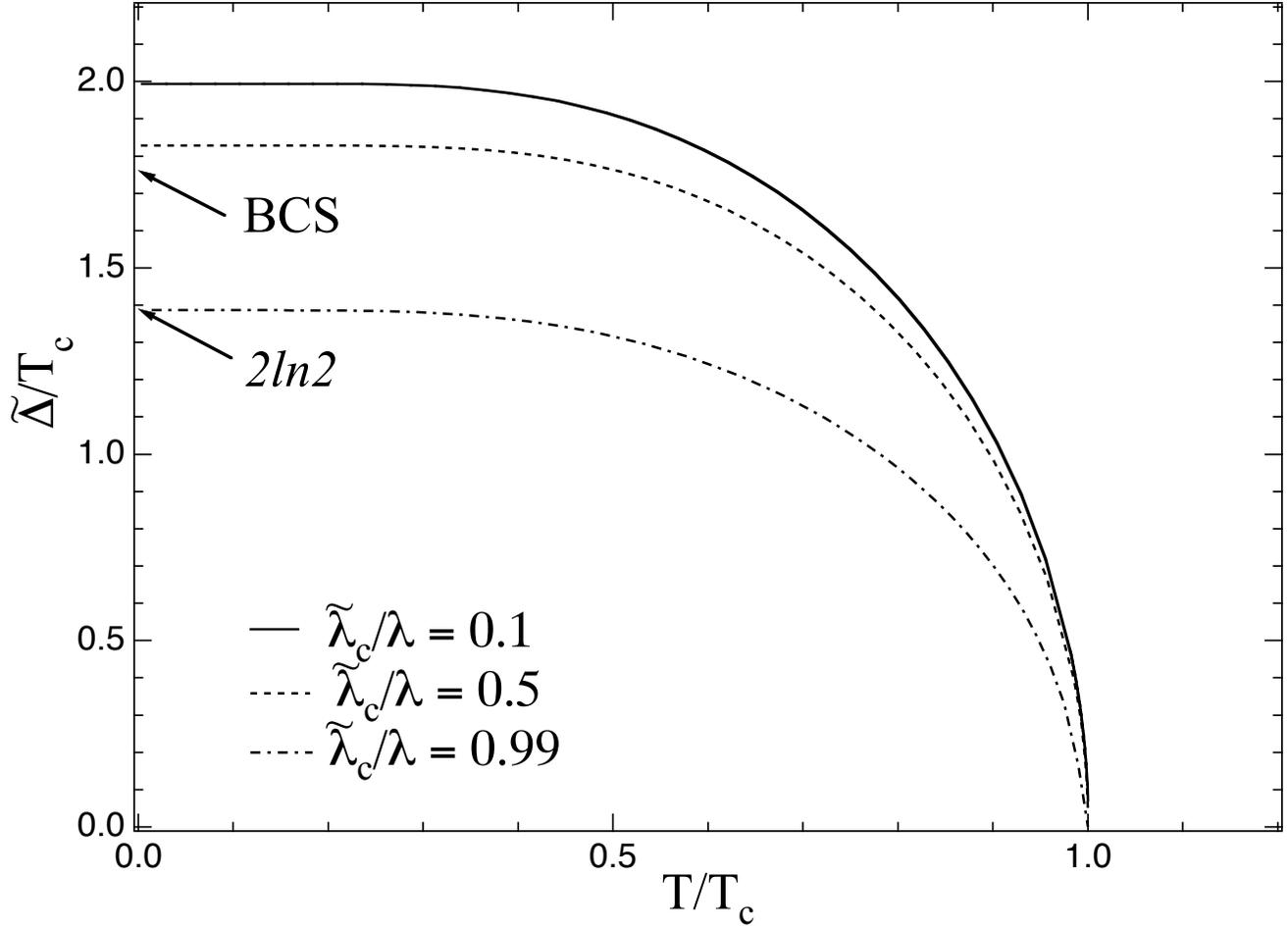} } \caption{The superconducting gap $
\tilde{\Delta}(T) $ divided by $ T_c $ as a function of the
normalized temperature $ T / T_c $ for several values of  $
\tilde{\lambda}_c / \lambda $. The two arrows indicate the values
for $ \tilde{\Delta}_0 / T_c $ given by the BCS theory and the $
\lambda  \simeq \tilde{\lambda}_c $ case ($\tilde{\lambda}_c=
\alpha/\Lambda$). } \label{fig4}
\end{figure}

Our results highlight the qualitative and quantitative differences
existing between Dirac electrons -- for which the Fermi surface
reduces to a point -- and quasi-free electrons, having a dispersion
relation $\epsilon(\vec k) = |\vec k|^2 / 2m^* $, which leads
to a Fermi surface formation. The properties of the former are
analyzed when the conditions are such that a superconducting
interaction is present in a quasi-two-dimensional system consisting
of a stack of $N$ planes. One of the most striking differences
between the two electron types, is the polynomial, rather than
exponential gap dependence on the inverse coupling constant, in the
case of Dirac electrons. This leads to a quantum phase transition
separating a normal from a superconducting phase for a critical
value of the coupling, at $T=0$. This transition is rather similar
to the one occurring in the Nonlinear Sigma model in 2+1D, where the
spin stiffness has an expression identical to (\ref{s00})
\cite{chn}. Because of this quantum phase transition, our model
reproduces qualitatively the superconducting transition in high-Tc
cuprates in the underdoped regime as one can infer from Fig. \ref{fig2}. This
seems to indicate that Dirac electrons may have an important role in
the mechanism of high-Tc superconductivity.

The ratio between the zero temperature gap to $T_c$, as well as the
behavior of the gap around $T_c$, are results also significantly
different for the case of Dirac electrons. It would be very
interesting to compare our results with experimental measurements of
these quantities in quasi-two-dimensional materials containing Dirac
electrons.

An interesting outcome of this work is the analysis the phase
structure generated by the model in the presence of a natural
physical cutoff, specially the regime where the cutoff is of the
same order of the gap and we are away from the quantum critical
point. In this case the (naturally) cutoff theory yields results
that are strongly different from the ones derived in the weak
coupling regime where the cutoff is much larger than the gap and the
system is close to the quantum critical point. The strong coupling
regime, in particular, is probably relevant for the high-Tc
cuprates.

\bigskip
\bigskip
\bigskip
We would like to thank S.Sachdev and A.H.Castro Neto, for very
helpful comments and conversations.

 This work has been
 supported in part by CNPq and FAPERJ. ECM has been
partially supported by CNPq. LHCMN has been supported by CNPq.

\bibliography{apssamp}

\begin{thebibliography}{400}


\bibitem{slee} S.H.Simon and P.A.Lee, Phys. Rev. Lett. 78 (1997)
1548; A.C.Durst and P.A.Lee, Phys. Rev. B 62 (2000) 1270
\bibitem{ferr} E.J.Ferrer, V.P.Gusynin and V. de la Incera,
Mod.Phys.Lett. B 16 (2002) 107
\bibitem{herbut} I.F.Herbut, Phys. Rev. Lett. 88 (2002) 047006
\bibitem{ft} M.Franz and Z.Te\u{s}anovi\'{c}, Phys. Rev. Lett. 84
(2000) 554, Phys. Rev. Lett. 87 (2001) 257003
\bibitem{and} P.W.Anderson, cond-mat/9812063 (unpublished)
\bibitem{sem} G.Semenoff, Phys. Rev. Lett. 53
(1984) 2449
\bibitem{hald} F.D.M.Haldane, Phys. Rev. Lett. 61 (1988) 2015
\bibitem{guinea} J.Gonzalez, F.Guinea and M.A.H.Vozmediano, Phys. Rev.
Lett.69 (1992) 172; Nucl. Phys. B406 (1993) 771; Nucl. Phys. B424
(1994) 595; Phys. Rev. Lett. 77 (1996) 3589; Phys. Rev. B 59 (1999)
2474; Phys. Rev. B 63 (2001) 134421
\bibitem{cn} N.M.R.Peres, F.Guinea and A.H.Castro Neto,
cond-mat/0506709
\bibitem{zhang} Y.Zhang et al., Nature 438 (2005) 201
\bibitem{nov} K.S.Novoselov et al., Nature 438 (2005) 197
\bibitem{castroneto} A.H.Castro Neto, Phys. Rev. Lett. 86 (2001)
4382
\bibitem{cnuch} B.Uchoa, A.H.Castro Neto and G.G.Cabrera, Phys. Rev. B
69 (2004) 144512; B.Uchoa, G.G.Cabrera and A.H.Castro Neto, Phys.
Rev. B 71 (2005) 184509
\bibitem{bfn} L.Balents, M.P.A.Fisher and C.Nayak, Int. J. Mod.
Phys. B 12 (1998) 1033
\bibitem{bf} L.Balents and M.P.A.Fisher, Phys. Rev. B 55 (1997) R
11973
\bibitem{mw} N.D.Mermin and H.Wagner,  Phys. Rev. Lett. 17 (1966)
1133; P.C.Hohenberg, Phys. Rev. 158 (1967) 383; S.Coleman, Commun.
Math. Phys. 31 (1973) 259
\bibitem{wl} X-G.Wen and P.A.Lee, Phys. Rev. Lett. 76 (1996) 503
\bibitem{mm} E.C.Marino and M.J.Martins, Phys. Rev. D 33 (1996) 3121
\bibitem{semenoff} G.W.Semenoff and L.C.R.Wijewardhana, Phys. Rev. Lett. 63
(1989) 2633
\bibitem{ros} B.Rosenstein, B.J.Warr and S.H.Park, Phys. Rev. Lett.
62 (1989) 1433
\bibitem{coleman} S.Coleman and E.Weinberg, Phys. Rev. D 7 (1973)
1888
\bibitem{oko} A.Okopi\'{n}ska, Phys. Rev. D38 (1988) 2507
\bibitem{rwp}B.Rosenstein, B.J.Warr and S.H.Park,  Phys. Rev. D39
(1989) 3088
\bibitem{appel} T.Appelquist and M.Schwetz, Phys. Lett. B491 (2000)
367
\bibitem{babaev} E.Babaev, Phys. Lett. B497 (2001) 323
\bibitem{witt} E.Witten, Nucl. Phys. B145 (1978) 110
\bibitem{kt} V.L.Berezinskii, Zh. Eksp. Teor. Fiz. 59 (1970) 907;
J.Kosterlitz and D.Thouless, J. Phys. C6 (1973) 1181
\bibitem{pg} V.M.Loktev, R.M.Quick and S.G.Sharapov, Phys. Rep. 349
(2001) 1
\bibitem{kv} D.V.Khveshchenko and H.Leal, Nucl. Phys. B687 [FS] (2004) 323
\bibitem{chn} S.Chakravarty, B.I.Halperin and D.Nelson, Phys. Rev. B 39
(1989) 2344
\end{thebibliography}

\newpage

\end{document}